\documentclass[amsmath,11pt,amssymb]{revtex4}

\usepackage{natbib}

\usepackage{color}
\usepackage{xcolor}
\usepackage[]{graphicx}
\usepackage{latexsym}
\usepackage{amsmath}
\usepackage[caption=false]{subfig}
\usepackage{multirow}
\usepackage{mathtools}
\usepackage{textcomp}
\usepackage{xr}
\usepackage{slashed}
\usepackage{dsfont}

\usepackage[pdftex,hyperfigures,breaklinks,colorlinks,citecolor=black,linkcolor=black]{hyperref} 

\newcommand{\be}{\begin{equation}}
\newcommand{\ee}{\end{equation}}

\begin{document}

\title{Predicting public market behavior from 
private equity deals}

\author{Paolo Barucca}
\affiliation{Department of Computer Science, UCL, London, UK}
\author{Flaviano Morone}
\affiliation{Center for Quantum Phenomena, Department of Physics, 
New York University, New York, NY 10003 USA}

\begin{abstract}
We process private equity transactions to predict 
public market behavior with a logit model. 
Specifically, we estimate our model to predict 
quarterly returns for both the broad market and for 
individual sectors. Our hypothesis is that private 
equity investments (in aggregate) carry predictive 
signal about publicly traded securities. The key source 
of such predictive signal is the fact that, during their 
diligence process, private equity fund managers are 
privy to valuable company information that may not 
yet be reflected in the public markets at the time 
of their investment. Thus, we posit that we can discover 
investors’ collective near-term insight via detailed 
analysis of the timing and nature of the deals they 
execute. 
We evaluate the accuracy of the estimated model 
by applying it to test data where we know the correct 
output value. Remarkably, our model performs consistently 
better than a null model simply based on return statistics, 
while showing a predictive accuracy of up to 70\% in 
sectors such as Consumer Services, Communications, 
and Non Energy Minerals.
\end{abstract}

\maketitle

\section{Introduction}

Many investment strategies are driven via complex 
representations of Big Data from diverse sources, 
such as social media, analyst reports, Bloomberg 
news, technical trading data, and macroeconomic 
data, by means of statistical or machine learning 
techniques. 

Our approach is different, in that it relies on 
actions taken by human experts to effectively 
filter this huge amount of data. We mine the 
actions of Private Equity (PE) investors after 
they have assimilated these large data sets: 
we leverage PE economic actors as 
{\it `human autoencoders'} providing an informative 
hidden representation of the original data. 
This approach naturally yields a low-dimensional 
statistical modeling problem that can be tackled 
with shallow and faster discriminative classifiers 
such as a logit model.

The basic premise of our modeling effort is that 
one of the factors that creates excess returns for 
private equity (over public equity) is the predictive 
ability of the large private equity investors (e.g., 
Bain Capital, KKR, BlackRock, Carlyle Group) with respect 
to which industries are likely to perform best. 
That is, we posit that a large portion of the reason why 
private equity fund managers continue to outperform market 
averages is their ability to identify those industries/sectors 
that will outperform others over a multi-year horizon.
This may be so not only because of such investors are privy 
to confidential information regarding the private companies 
in which they are considering investments, but also because 
of their expectations in private equity markets, contract 
commitments, aggressive approaches with a goal to affect 
competitors, or by imposing skilled management on a private 
company they purchase, buying the company at a favorable price, 
etc.

In the bigger picture, our hypothesis leverages the 
theory of rational expectations proposed by Muth in 
1961~\cite{muth61}, according to which agents make 
decisions based on all information available to them 
combined with past experiences. Therefore, expectations 
of investors, since they are informed forecasts of future 
outcomes, tend to be the same as the theoretical predictions 
of the underlying economic model. 
Simply put, we suggest that such investments by PE fund managers 
may be a self-fulfilling prophecy, in that the prices of public 
companies, like those of the private companies in which they have 
invested in the rational expectation of gain, may experience a 
boost (rather than a fall).

\section{Hypothesis statement}
Unlike public companies, which must publicly release 
financial statements and any information that is materially 
relevant for shareholders, private firms in the U.S. generally 
face no such requirements. Therefore, private firms may reveal 
information to prospective investors as long as there is no 
misrepresentation or omission of material information in connection 
with the sale or purchase of a security (which would be at odds 
with SEC Rule 10b-5), although there is no requirement that such 
disclosures must be publicly made. 
By virtue of this we would like 
to suggest that private equity investments carry predictive 
signal about publicly traded instruments, such as equities, 
indices, etc. 
This hypothesis stems from the fact that private equity investors 
are privy to confidential information provided to 
them by private companies in which they are 
considering investments~\cite{fenn1997private, connelly2011signaling}. Such information may be 
both systematic, i.e. relates to general market value, 
or specific, i.e. relates to both the individual 
company and to its market 
sector~\cite{folta2004strategic,janney2003signaling}. 
The investor's 
assessment of such confidential information (but 
not the information itself) is revealed publicly 
when it makes an initial investment in the company. 
We posit that such assessment carries predictive 
value with respect to publicly traded proxies 
for those investments (e.g. equities of firms in 
the same sector). In other words, we posit that, 
when a PE investor initiates a new investment in 
a private company, its positive opinion of that 
investment is informed by the investors’ access 
to privileged information that is not already 
priced in by the public market. 

We base our use of first-deal data (i.e., the 
first investment in a given company) on the fact 
that these investments are the most bias-free, 
i.e. the investor affirmatively elected to make 
an investment at that time in that sector. So, 
we may well expect these first deals to have the 
most predictive value.
The act of making each first investment is quickly 
made available to us by a commercial information 
service provider such as FactSet~\cite{FactSet}, 
who monitors and publishes such information in a 
timely fashion.
Furthermore, that first deal information is enriched 
with additional characteristics of the investor and 
investment, which can be used to inform a predictive 
model and enhance predictive success, as explained 
in Section IV. 

We have access to historical data on all significant 
PE deals, and we can quantitatively assess the 
relationship between PE investments and publicly 
traded equities. 
The investors’ market timing may reveal their 
aggregate, informed opinion about changes in the 
near-term value of public stocks. Likewise, 
investors’ choice of market sector may reveal 
their informed opinion about relative sector behavior. 
In this work, we therefore elect to model both the 
systematic public market behavior (as represented by 
the FactSet US Index) and specific public market 
behavior (as represented by the 19 FactSet sector 
indices that comprise the FactSet US Index).

\section{Literature Review}
Quantifying how prices adjust to new information 
is central in finance~\cite{fama1970efficient} and 
it is the subject of the theory of rational 
expectations~\cite{muth61} and of signaling 
theory~\cite{janney2003signaling}.
In private equity investments many actors with different 
information are involved: the issuers, the companies which 
need this form of financing to foster their rapid growth, 
the intermediaries, if any, who participate in the 
investment management, perform -directly or indirectly through agents- 
fundamental research on the issuers, and ultimately help assessing the risk-return spectrum for the investment; and the investors, who constitute the 
money supply looking for medium- and long-term financial deals. 
The economic literature has long been investigating how rational expectations can translate into price movements~\cite{muth61}, and in signaling theory~\cite{connelly2011signaling} how such expectations maps the information into equity placements. For a recent summary of the available evidence on the applications of signaling theory see~\cite{svetek2022signaling}.
Private equity placements have been found to be associated with significant 
price discounts and yet lead to positive returns~\cite{wruck1989equity, hertzel1993market}, reflecting a form of resolution of asymmetric information 
concerning the company valuation. 
In particular, the work on signaling~\cite{janney2003signaling} has shown 
how private equity placements can affect the  valuation of firms specifically in the biotechnology domain. It has been found that firms, perceived 
as uncertain by the market, tend to be able to  attract more financing, if they are able to place their private equity and later disclose their 
investors. 

Information acquisition is costly, so investors that are willing to analyse and buy private equity from a company can play a major role into its valuation, in particular if these investments happen at an early stage of the corporate life cycle. 
Investment ties can play a fundamental role in the success of a company depending on its corporate life cycle and on other market 
conditions~\cite{gulati2003ties}. Investors' experience and education strongly affects first-round financing~\cite{ko2018signaling}, as investors with an high-profile or a pivotal role in the industry can influence the expectations of follow-on investors and less experienced investors that cannot rely on first-hand information, as observed both 
in~\cite{capizzi2022business} and~\cite{drover2017attributes} with the role of angel and crowdfunded investments, 
in~\cite{bernstein2017attracting} which considered the role of the founding team, and in~\cite{croce2017business} studying the role of venture capitalists. This positive effect may have its negative counterpart, i.e. the divestment or withdrawal of central investors may affect negatively the valuation of a company and its 
funding~\cite{shafi2020investment}. Recently, it was also observed how media attention can help firms in signaling their unrealised performance~\cite{vanacker2020signal}.

Recent studies have also looked at the relationship between market discounts of private equity placements and their subsequent market performance: in~\cite{chen2010earnings} the authors investigate how pre-issuance firm earnings statements can be related to a firm's private equity placement strategy, in~\cite{Liang2013InformationAA} the authors suggest that the market discounts serve as a compensation for investors' costs of assessing companies, and how growth firms and value firms can react differently to private equity placement~\cite{Chuang2020PrivatePM}.

Private equity placements can lead to long-term benefits to firms 
by balancing the information asymmetry between the firms and the investors~\cite{folta2004strategic}, and private investors become a factor that differentiates the firm from others in the same sector~\cite{janney2006moderating}.
This kind of signaling, which allows firms to 
communicate an undervaluation by seeking private investments, ultimately balances information 
asymmetry~\cite{connelly2011signaling,Liang2013InformationAA} in financial 
markets, thus tackling directly market inefficiencies. 

Despite the evidence on the general role of private equity 
in direct signaling for firms' valuation, under multiple conditions and considering multiple factors, no study has recently investigated the indirect effect of private equity placements in financial markets across different sectors. 
Our study contributes to the overall body of evidence on the role of informed private investments in signaling market returns, in particular by showing evidence of cross-signaling at a sector level: private investments in a given sector carry signal over the market performance of the public stocks in the corresponding sector. 
The contribution of this study is to consider both a universal 
and multi-sector logit model~\cite{mcfadden1973conditional, greene2003econometric} to 
link the cross-dependence of positive public market's forward 
returns on the volume of private equity placements across sectors.

\section{Methodology}
Our modeling approach requires five essential 
steps including: 
(1) Data Extraction; 
(2) Feature Preparation; 
(3) Standardization of the features; 
(4) Construction of the Response variable; 
(5) Model Definition and Parameters' Estimation.

\subsection{Data Extraction}
\begin{figure}[h!]
\includegraphics[width=\textwidth]{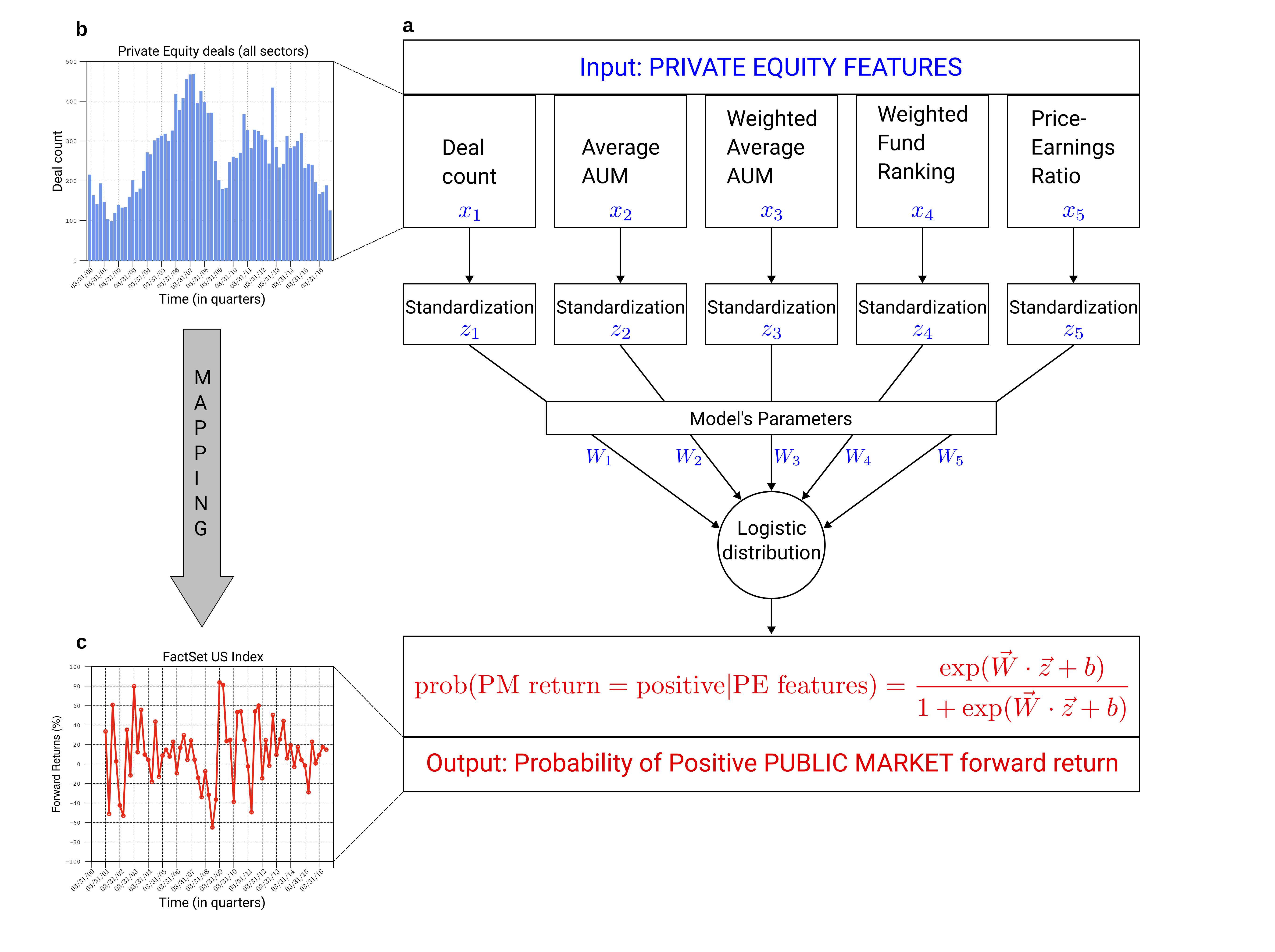} 
\caption{{\bf Logit model to infer the mapping 
between Private Equity deals and Public Market 
positive forward returns.} 
{\bf a}. The model takes the input of different PE 
features and store the input values in variables $x_i$. Then, these variables are standardized using Eq.~\eqref{eq:zscore} into z-scores $z_i$. 
The vector of z-scores is combined with the parameters of 
the logit model and converted into a probability 
via the logistic distribution function. 
This probability is the prediction of the model 
regarding the positive forward return of the Public Market 
(the FactSet US Index or the sectors therein contained). 
{\bf b}. Total number of PE deals per quarter 
in all market sectors representing an example 
of PE feature fed as input to our model. 
The model must be estimated over a certain number 
of quarters in order to learn the mapping, i.e. 
the hidden relationship, between PE investments 
and Public Market (PM) returns. 
{\bf c}. Forward returns of the FactSet Us Index
representing the response variable of our model. 
During the estimation phase many samples 
consisting of pairs {\it $\langle$PE features, 
PM return$\rangle$} are `shown' to the model, 
which estimates the parameters by maximizing 
the likelihood function defined in Eq.~\eqref{eq:likelihood}. 
}
\label{fig:Model}
\end{figure}

The first step consists in extracting private equity 
investment data from FactSet. We are interested in 
data of completed transactions (i.e. buyout and growth 
expansion) in North America. 
Thus, the source data used in our study was extracted from 
FactSet~\cite{FactSet} based on filtering rules and 
consists of data on: 
\begin{itemize}
\item Portfolio Companies (unique records) 
\item Investors 
\item Investment Rounds
\end{itemize}
Based on our filtering rules, we extracted from FactSet: 
\begin{itemize}
\item 16,668 unique Portfolio Company records from Jan 2000 to Dec 2016
\item 31,505 Investor records
\end{itemize}
Since, in general, there may be multiple records for the 
same portfolio company (that is, unique combinations of 
Investor, Portfolio Company, and Deal), we further 
filtered them down, using certain investment type 
criteria. Specifically, using the investor data, we 
identified the date of the~\underline{first} 
investment (deal) in each portfolio company (positing 
that this data provides us the earliest indicators 
of private equity investment trends), as shown in Table~\ref{tab:table1x}. 
We then aggregate the counts of such investments by: 
\begin{itemize}
\item Quarter
\item Sector 
\end{itemize}
The act of making each first investment reveals to us 
the investors’ opinions with respect to both market 
timing (the ``when'', given by the date of the 
{\it First Investment} tab in Table~\ref{tab:table1x}) 
and selection of sector (the ``what'', given by 
the {\it FactSet Sector} tab in the same table). 
Further, that first-deal information is enriched 
with additional characteristics of the investor and investment, 
which, instead, will be used in our predictive model. 
The size and type of investor and investment can 
play a crucial role in the perception of the signal 
from the market. In particular, we have data about Investor 
AUM (asset under management) and Investor's Performance 
(in the form of a quartile ranking). 
An excerpt from the full data set is shown in Table~\ref{tab:table1x}:
these variables represent the raw features to be 
fed as input into the logit model depicted in Fig.~\ref{fig:Model}a. 
\begin{table}[h!]
\begin{center}
\begin{tabular}{|c|c|c|c|c|}
\hline
Company Name & FactSet Sector  & First Investment & Investor AUM  & Investor Performance\\
\hline
21st Century Oncology Holdings, Inc. & Health Services & Feb-12-08 & 
AUM $>10$ & N/A\\
\hline	
JW Resources, Inc. & Energy Minerals & Feb-01-13 & 
$2<$AUM$<10$ & Bottom Two Quartiles\\
\hline
ACE Cash Express, Inc. & Finance & Oct-06-06 & 
$2<$AUM$<10$ & Top Two Quartiles\\
\hline
$\dots$ & $\dots$ & $\dots$ & $\dots$ & $\dots$\\
\hline
\end{tabular}
\caption{Excerpt of first-deal data and Investors's 
characteristics export from FactSet. We filtered all 
companies whose status is private. We have also 
information about Investor's AUM (in Billions of US\$) 
and Investor's Performance. 
N/A means that the given field is not populated 
with that specific investor data. 
}
\label{tab:table1x}
\end{center}
\end{table}

\subsection{Feature Preparation}
We build our model at two levels of 
granularity: {\it (i)} at the level of the 
FactSet US Index to predict the future behavior 
of the broad US market; {\it (ii)} at the level 
of the 19 sector indices that comprise the FactSet 
US Index to predict future sector behavior, 
as explained next. 

\subsubsection{FactSet US Index Model}
The following features have been included as 
input variables to the logit model of the 
FactSet US Index: 
\begin{enumerate}
\item {\bf Deal count}. This is defined as the total 
number of PE deals (i.e. first round investment) in 
the given quarter, as seen in Fig.~\ref{fig:Model}b.  
Among all possible PE features, first-deal counts 
is the most rational choice consistent with our objective of 
predicting public market forward return. Indeed, our initial 
premise is that the number of investments increases (in all sectors, 
or in one sector) in advance of a positive change in the 
public market returns. So, we expect these first deals to 
have the most predictive value.
\item {\bf Average AUM}. 
This is defined as the total AUM of investors closing 
PE deals in the given quarter divided by the number of 
first-deal counts, i.e.
\begin{equation}
{\rm Avg\ AUM} = \frac{\rm Total\ AUM}{\rm Deal\ count}\ .
\end{equation} 
Intuitively, the choice of the AUM as a predictive feature 
stems from the fact that if the AUM of all investors in a 
given quarter is large, then we may anticipate a gain in 
the forward return of the broad market. 
\item {\bf Weighted Average AUM}.  
Alongside the simple AUM, it may be also meaningful 
to create a different version of this feature, by 
assuming that private investments made by large investors 
(i.e. AUM $> \$10$B) are a stronger indicator of 
a positive change in the public market return in the 
next quarter than investments made by smaller investors 
(i.e. AUM $< \$2$B). 
The Weighted Average AUM feature is designed precisely 
to reward more those deals made by large investors rather 
than deals made by small investors. 
The precise value of the weights is not particularly important 
as long as large investors are assigned bigger weights than 
small investors. Therefore, we define the Weighted Average AUM 
as the average AUM weighted as follows
\begin{equation}
{\rm Weight} = 
\left\{
\begin{matrix}
0.1 & {\rm if\ } & {\rm AUM} < \$2B\\
0.5 & {\rm if\ } & {\rm AUM} \in [\$2B,\$10B]\\
1.5 & {\rm if\ } & {\rm AUM} > \$10B
\end{matrix}
\right.\ ,
\label{eq:weights}
\end{equation} 
where the weights are chosen solely with the goal 
to magnify the difference between small and large 
investors, and thus are by no means fine-tuned to 
any particular optimal value (if any). 
\item {\bf Average Fund Ranking}. This is defined as 
the average quartile ranking computed by averaging 
the performance rankings over all sectors comprising 
the FactSet US index. 
This ancillary feature quantifies investors' quality 
and we may expect that deals made by investors ranked in 
the top tier to have higher success in predicting positive 
forward return in the public market than  deals made by 
investors ranked in the bottom tier. 
It is important to make the following remark 
about the values of this feature. 
That is, by aggregating the quartile rankings from all 
sectors we obtained a sufficiently smooth time series 
without missing data that can be used in our broad 
market model. However, due to the fact that for 
some sectors this field was lightly populated (or 
data were missing altogether, as seen in Table~\ref{tab:table1x}) 
the Average Fund Ranking has not been used as input 
feature in the Sectors Model.
\item {\bf Price-Earnings Ratio}. This is defined as 
the average Price divided by the Earnings for all stocks 
in the FactSet US Index in the given quarter. 
Although this is not a `private equity feature', we include 
it in the model as well. 
The inclusion of this additional exogenous information 
from the public market may provide us with additional 
model refinement, because it characterizes somehow the 
investment environment and its potential for 
suppressing/enhancing the number of first deals.
\end{enumerate}

An example of feature values used as input to the 
FactSet US Index model for several quarters is 
shown in Table~\ref{tab:table3}. 
\begin{table}[h!]
\begin{center}
\begin{tabular}{|c|c|c|c|c|c|c|}
\hline
Sector & Date & Deal Count &Avg-AUM & Weight-Avg-AUM & 
Avg-Fund-Ranking & PE ratio\\
\hline
FactSet US & 03/31/08 & 397 & 6.79 & 4.55 & 2.42 & 18.45\\
\hline 
FactSet US & 06/30/08 &	369 & 5.67 & 3.70 & 2.29 & 18.53\\
\hline	
FactSet US & 09/30/08 &	370 & 5.61 & 4.15 & 2.45 & 17.45\\
\hline	
\hline
FactSet US & 12/31/08 &	248 & 3.93 & 3.98 & 2.26 &	14.34\\
\hline	
FactSet US & 03/31/09 &	200 & 2.29 & 2.82 & 2.41 &	14.38\\
\hline
FactSet US & 06/30/09 &	178 & 6.08 & 4.10 & 2.46 &	16.94\\
\hline
FactSet US & 09/30/09 &	181 & 4.35 & 3.59 & 2.14 &	19.30\\
\hline
\end{tabular}
\caption{Features of the FactSet US Index model 
for several quarters. The double line marks the 
date of the financial crisis followed by the Great Recession. 
Notice how in the year 2009 the number of PE deals in a given 
quarter is about half of that in the corresponding quarter of 
the year before. (AUM in Billions of US\$.)}
\label{tab:table3}
\end{center}
\end{table}

\subsubsection{Sectors Model}
The 19 sectors comprising the FactSet Us Index 
are: 1) Commercial Services;
2) Communications;
3) Consumer Durables;
4) Consumer Non-Durables;
5) Consumer Services;
6) Distribution Services;
7) Electronic Technology;
8) Energy Minerals;
9) Finance;
10) Health Services;
11) Health Technology;
12) Industrial Services;
13) Non-Energy Minerals;
14) Process Industries;
15) Producer Manufacturing;
16) Retail Trade;
17) Technology Services;
18) Transportation;
19) Utilities.

The following features have been included 
as input variables to the estimated logit model 
of each of the 19 sectors listed above. 
\begin{enumerate}
\item {\bf Sector Deal count}: total number of 
deals in PE companies belonging to the given 
sector in the given quarter. 
\item {\bf Sector counts as a percentage of total}. 
This is defined as
\begin{equation}
{\rm Sector\ Count\ as\ \%\ of\ Total} = 
\frac{\rm Total\ Count\ for\ the\ Sector}
{\rm Total\ Count\ for\ all\ Sectors}\ .
\end{equation}
This version of the Sector Deal Count feature is 
consistent with our objective of predicting relative behavior 
(i.e., of a sector to a broad index).
\item {\bf Average AUM} of investors closing 
PE deals in the given sector in the given quarter, 
defined as 
\begin{equation}
{\rm Avg\ Sector\ AUM} = \frac{\rm Total\ Sector\ AUM}{\rm Sector\ Deal\ count}\ .
\end{equation} 
\item {\bf Sector Weighted Average AUM}: average 
weighted AUM for the given sector, computed by 
using the same weights as in Eq.~\eqref{eq:weights}. 
\item {\bf Sector Price-Earnings Ratio}: average price 
divided by the Earnings for all stocks in the given 
sector.
\item {\bf FactSet Price-Earnings Ratio}: average 
price divided by the Earnings for all stocks in the 
FactSet US Index. (Although this is not a sector 
specific feature, we include it in the sector model 
as well.) 
\end{enumerate}
An example of feature values used as input to the 
Sector model for the {\it Commercial Services} 
sector is shown in Table~\ref{tab:table4}. 
\begin{table}[h!]
\begin{center}
\begin{tabular}{|c|c|c|c|c|c|c|c|}
\hline
Sector & Date & Deal Count & Count as \% of Tot &
Avg-AUM & Weight-Avg-AUM & Sector PE & 
FactSet PE\\
\hline
Commercial Services & 03/31/08 & 49 & 16.28 & 5.07 &
4.24 & 19.57 & 18.45\\
\hline
Commercial Services & 06/30/08 & 38 & 13.92 & 4.40 & 
2.83 & 19.19 & 18.53\\
\hline
Commercial Services & 09/30/08 & 37 & 12.80 & 4.50 &
2.55 & 17.92 & 17.45\\
\hline	
\hline 
Commercial Services & 12/31/08 & 24 & 12.06 & 3.80 &
3.96 & 14.11 & 14.34\\
\hline
Commercial Services & 03/31/09 & 22 & 13.33 & 1.09 & 
2.15 & 13.28 & 14.38\\
\hline
Commercial Services & 06/30/09 & 21 & 13.82 & 1.56 & 
3.00 & 18.02 & 16.94\\
\hline
Commercial Services & 09/30/09 & 20 & 13.70 & 1.51 & 
2.71 & 20.62 & 19.30\\
\hline
\end{tabular}
\caption{Features of the Sector model for 
the {\it Commercial Services} sector. 
(AUM in Billions of US\$.)}
\label{tab:table4}
\end{center}
\end{table}

\subsection{Standardization of the features}
\begin{figure}[h!]
\includegraphics[width=0.95\textwidth]{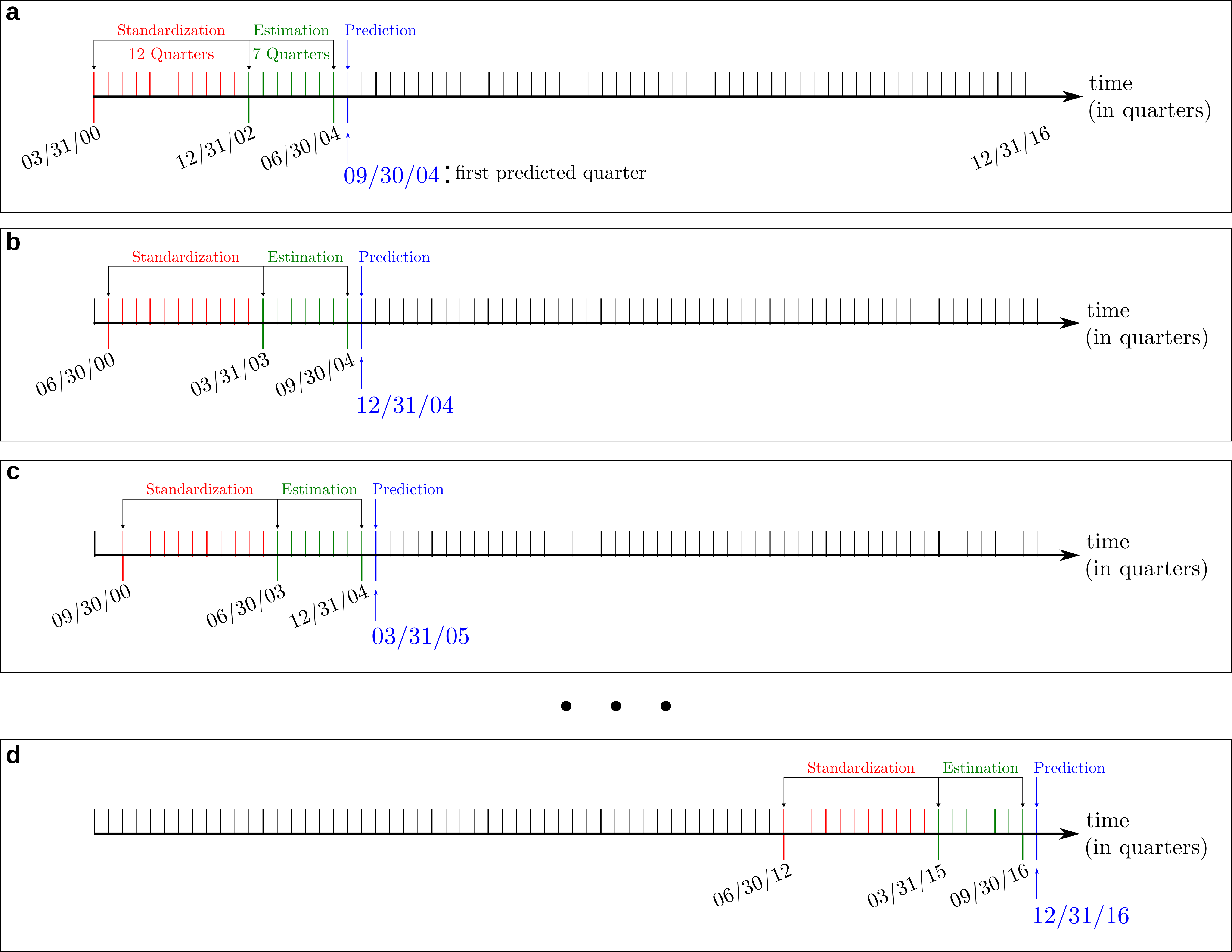} 
\caption{{\bf Standardization, estimation, and 
quarterly prediction process.} 
{\bf a}, We use $T=12$ quarters to standardize the 
feature vector (i.e. to compute the z-scores), 
shown in red; then we estimate the model using 
$N_E=7$ quarters, shown in green; 
finally, we predict the return of the next 
quarter, shown in blue. In this setting the first 
predicted quarter falls on the date $09/30/04$. 
{\bf b}, After that, we roll down the black arrows 
by one quarter, recompute the z-score, 
and re-estimate the model's parameters to predict 
the next quarter dated $12/31/04$. More precisely, 
we use a rolling window approach, where the model is 
re-estimated using a 7-quarters period, including the 
09/30/04 sample, and a one-ahead forecasting is conducted 
for the $12/31/04$ quarter. 
{\bf c}, Similarly to {\bf b}, we re-estimate the 
model using a 7-quarters period, including the 
12/31/04 sample, and make a prediction for the subsequent 
quarter dated $03/31/05$.
{\bf d}, We keep sliding forward the 7-quarters estimation 
window, and corresponding one-ahead forecast, 1 quarter at time until 
the last quarter of our data set has been predicted, dated $12/31/16$.
}
\label{fig:training}
\end{figure}

The rationale behind the normalization of the 
features values is to smooth the time series data, 
particularly where they are low and erratic, by using 
a simple metric of z-scores relative to an average of 
the last 3 years (i.e., 12 quarters). 
It suffices to explain the procedure for a 
single feature. Let us call $x(t)$ the value of 
feature $x$ (for example deal count) at time $t$. 
Time is discretized in quarterly intervals, so in 
one year there are $4$ time points. We have in total 
$N_q=68$ quarters, corresponding to $17$ years of 
data from 03/31/2000 to 12/31/2016. 
To normalize the features we define a time window 
$T$ and compute the mean and variance of $x(t)$ 
in this time window as 
\begin{equation}
\begin{aligned}
\mu(t) &=
\frac{1}{T}\sum_{\tau=t-T+1}^{t}x(\tau)\ ,\\
\sigma^2(t) &=
\frac{1}{T-1}\sum_{\tau=t-T+1}^{t}[x(\tau)-\mu(t)]^2\ ,
\end{aligned}
\end{equation}
and then construct the z-score of feature $x$ as 
\begin{equation}
z(t) = \frac{x(t) - \mu(t)}{\sigma(t)}\ .
\label{eq:zscore}
\end{equation} 
In practice we choose $T=12$ quarters, so the 
first z-score is assigned on $12/31/2002$ 
as shown in Fig.~\ref{fig:training}.

\subsection{Construction of the Response variable}
We discuss here the construction of the `response' variable, 
i.e. what is to be predicted. 
Here we wish to predict the sign of the return 
of the public market one quarter forward, given 
the values of the PE features in the current 
quarter. 
In other words, we want to predict whether the 
public market, represented here by the FactSet 
US Index, will be UP or DOWN in the next 
quarter given our knowledge of the PE features 
in the present quarter. 
It is important to remark that we are not 
attempting to forecast the exact value of 
the forward return, but only its polarity, 
i.e. the probability of positive (UP) or 
negative (DOWN) returns.   
Thus, we choose the {\bf response variable} 
$y(t)$ for the FactSet US Index model to be 
the sign of the annualized 1-quarter forward 
return, defined as
\begin{equation}
y(t)\equiv {\rm sign}\big[R_{\rm AnnForw}(t)\big]=
{\rm sign}\left[\left(\frac{P(t+1)}{P(t)}\right)^4-1\right]\ ,
\label{eq:response_FactSet}
\end{equation} 
where $P(t)$ is the value of the Index at 
quarter $t$. Table~\ref{tab:table5} shows 
an example of calculation of 1-quarter 
forward returns, both simple and annualized, 
and the value of the response variable 
$y(t)$. For example, the entry on date $03/31/03$ is 
the annualized 1 quarter forward return 
relative to the period starting at $03/31/03$ 
and ending at $06/30/03$. 
In Fig.~\ref{fig:Model} c we show the annualized 
forward return of the FactSet US Index (broad market) 
for the whole dataset analyzed here. 
\begin{table}[h!]
\begin{center}
\begin{tabular}{|c|c|c|c|c|}
\hline
Date & FactSet Index Value & 1-Q forward return (\%) & 
Ann 1Q forw return (\%) & Response Var $y(t)$
\\
\hline
12/31/2002 & 66.43170 & $-3.04$ & $-11.60$ & -1\\
\hline
03/31/2003 & 64.41500 & 15.82 & 79.97 & 1\\
\hline
06/30/2003 & 74.60800 & ... & ... & ... \\
\hline
\end{tabular}
\caption{Example of calculation of annualized 
1-quarter forward returns and corresponding 
response variables. We note, {\it en-passant}, 
that we could have defined the response variable 
as the sign of the forward return, instead of 
the annualized forward return, since both 
have the same sign. 
}
\label{tab:table5}
\end{center}
\end{table}

\subsubsection{Response variable for the Sector model}
In the Sector model we wish to forecast 
the excess return of the sector relative 
to the broad market return. Precisely, 
we want to predict the sign of the excess 
return. 
Thus we define the {\bf Sector spread 1-quarter forward}
as the sector forward return (annualized) less the FactSet return:
\begin{equation}
{\rm Sector\ spread\ 1Q\ forward}(t) = 
R^{Sector}_{\rm AnnForw}(t) - R^{FactSet}_{\rm AnnForw}(t)\ ,
\end{equation}
whereby we can construct the {\bf sector response variable} as
\begin{equation}
y(t)\equiv {\rm sign}\big[{\rm Sector\ spread\ 1Q\ forward}(t)\big]\ .
\label{eq:response_Sector}
\end{equation} 
Table~\ref{tab:table6} shows an example of calculation 
of Sector spread 1-quarter forward for the 
{\it Commercial Services} sector and the value of the
corresponding response variable defined in 
Eq.~\eqref{eq:response_Sector}.
\begin{table}[h!]
\begin{center}
\begin{tabular}{|c|c|c|c|c|}
\hline
Date & Sector Ret 1Q Forw (\%) & FactSet Ret 1Q Forw (\%) & Sector Spread 1Q Forw (\%)& Resp Var $y(t)$
\\
\hline
12/31/2002 & $-17.90$ & $-11.60$ & $-6.30$ & -1\\
\hline
03/31/2003 & $124.67$ & $79.97$  & $44.70$ & 1\\
\hline
06/30/2003 & ... & & & \\
\hline
\end{tabular}
\caption{Example of calculation of Sector spread 
1-quarter forward and of sector response variable 
for the {\it Commercial Services} sector.}
\label{tab:table6}
\end{center}
\end{table}

\subsection{Model Definition  and Parameters Estimation}
We use a logit model that attempts to forecast 
future positive quarterly returns of the public market 
from the knowledge of the volume and nature of 
private equity transactions. 
Specifically, using many samples consisting of PE features 
and public market returns we estimate a logit model to 
infer the `mapping' from the input space (PE transactions) 
to the output space (public market returns) 
as explained next. 

We consider a vector with $M$ features 
$\vec{z}=(z_1,\dots, z_M)$, which we want 
to classify in one of two possible classes. 
For example, for the FactSet US model, we 
can think of these features as the standardized 
values of: {\it (i)} Deal count, 
{\it (ii)} Average AUM, {\it (iii)} Weighted 
Average AUM, {\it (iv)} Average Fund Ranking, 
{\it (v)} Price-Earnings Ratio at a given 
quarter in time, as shown in Fig.~\ref{fig:Model}a. 
Hence in this case $M=5$. 
We label the two classes `UP' and `DOWN' in 
that they model the instances of positive 
and negative market returns, respectively. 
The quantity we wish to forecast is the probability 
of a positive forward return of the public 
market given the current values of PE features, 
denoted as $P({\rm UP}|\vec{z})$. 
The opposite case, i.e. the probability 
of a negative return $P({\rm DOWN}|\vec{z})$ 
can be simply obtained as 
\begin{equation}
P({\rm DOWN}|\vec{z}) = 1 - P({\rm UP}|\vec{z})\ , 
\end{equation}
since probabilities must add up to $1$. 
Therefore, it suffices to model, and compute, 
just one of the two probabilities, say 
$P({\rm UP}|\vec{z})$. 
We model $P({\rm UP}|\vec{z})$ via a logistic 
distribution function as follows
\begin{equation}
P({\rm UP}|\vec{z},\vec{W}, b) = 
\frac{\exp\left(\sum_{i=1}^M W_iz_i + b\right)}
{1 + \exp\left(\sum_{i=1}^M W_iz_i + b\right)}\ ,
\label{eq:pUP}
\end{equation}
where $\vec{W}=(W_1,\dots, W_M)$ and $b$ are 
parameters to be estimated. 
The set of parameters $(\vec{W}, b)$ reflects 
the impact of changes in the feature vector 
$\vec{z}$ on the probability of class UP.  
The goal is to estimate the model's parameters in order to 
compute the probability in Eq.~\eqref{eq:pUP} and, ultimately, 
make our predictions. 

To estimate the logit model defined in Eq.~\eqref{eq:pUP} we 
consider $N_E$ samples, each consisting 
of a pair $(\vec{z}(t), y(t))$ where both 
the feature vector and the response variable 
are known. 
These samples are fed to the logit model which 
infers the mapping of input space onto the output 
space by maximizing the `likelihood' function~\cite{mackay2003information,greene2003econometric}, defined as
\begin{equation}
L(\vec{W},b) = 
\sum_{t=1}^{N_E}\mathds{1}[y(t)={\rm UP}]\log P({\rm UP}|\vec{z},\vec{W}, b) + 
\mathds{1}[y(t)={\rm DOWN}]\log\big[1 - P({\rm UP}|\vec{z},\vec{W}, b)]\ ,
\label{eq:likelihood}
\end{equation}
with respect to $\vec{W}$ and $b$ (here $\mathds{1}[y(t)={\rm UP}]$ 
is equal to one if $y(t)={\rm UP}$ and zero otherwise). 
To maximize the likelihood we use a simple 
gradient-ascent algorithm whereby we update 
the model's parameters according to the following 
rules until convergence:
\be
\begin{aligned}
W_i^{(n+1)} &= W_i^{(n)} + 
\eta\frac{\partial L\big(\vec{W}^{(n)},b^{(n)}\big)}
{\partial W_i},\\
b^{(n+1)} &= b^{(n)} + 
\eta\frac{\partial L\big(\vec{W}^{(n)},b^{(n)}\big)}
{\partial b}, 
\end{aligned}
\ee
where $\eta$ is a small parameter that we set to $\eta=10^{-3}$ in our numerical experiments. 
We use $N_E=7$ samples to estimate the logit model. 
These are chosen to be $7$ consecutive quarters; 
then, having estimated the model's parameters,
we predict the return of the $(N_E+1)=8^{\rm th}$ quarter 
according to the following criterion:
\begin{equation}
{\rm Prediction\ of\ Public\ Market\ Return} = 
\left\{
\begin{matrix}
{\rm UP} & {\rm if}\ \ 
P({\rm UP}|\vec{z}(N_E+1)) \geq 0.5\ \\
{\rm DOWN} & {\rm if}\ \ 
P({\rm UP}|\vec{z}(N_E+1)) < 0.5\ 
\end{matrix}
\right.\ ,
\end{equation}
as shown in Fig.~\ref{fig:training}. 
Thus, the return of a given quarter is predicted 
based on the model's parameters estimated in the 
previous $7$ quarters.
The first quarter in our dataset is 
dated $03/31/00$; since we use $12$ quarters to 
normalize the data and $7$ more to estimate the model, 
then our first predicted quarter falls on date $09/30/04$. 
After that, we make a prediction for every 
subsequent quarter, in the rolling fashion 
illustrated in Fig.~\ref{fig:training}, thus 
collecting a total of 50 predictions for the 
FactSet US Index and 50 predictions for each 
of the 19 market sectors. 
Next we discuss the performance of our logit 
model by measuring the accuracy of our predictions 
both for the broad market and individual sectors.

\section{Numerical results}
In this section, we analyze the predictions of 
our logit model and provide evidence of private 
investment signaling. 

\begin{figure}[h]
\includegraphics[width=0.8\textwidth]{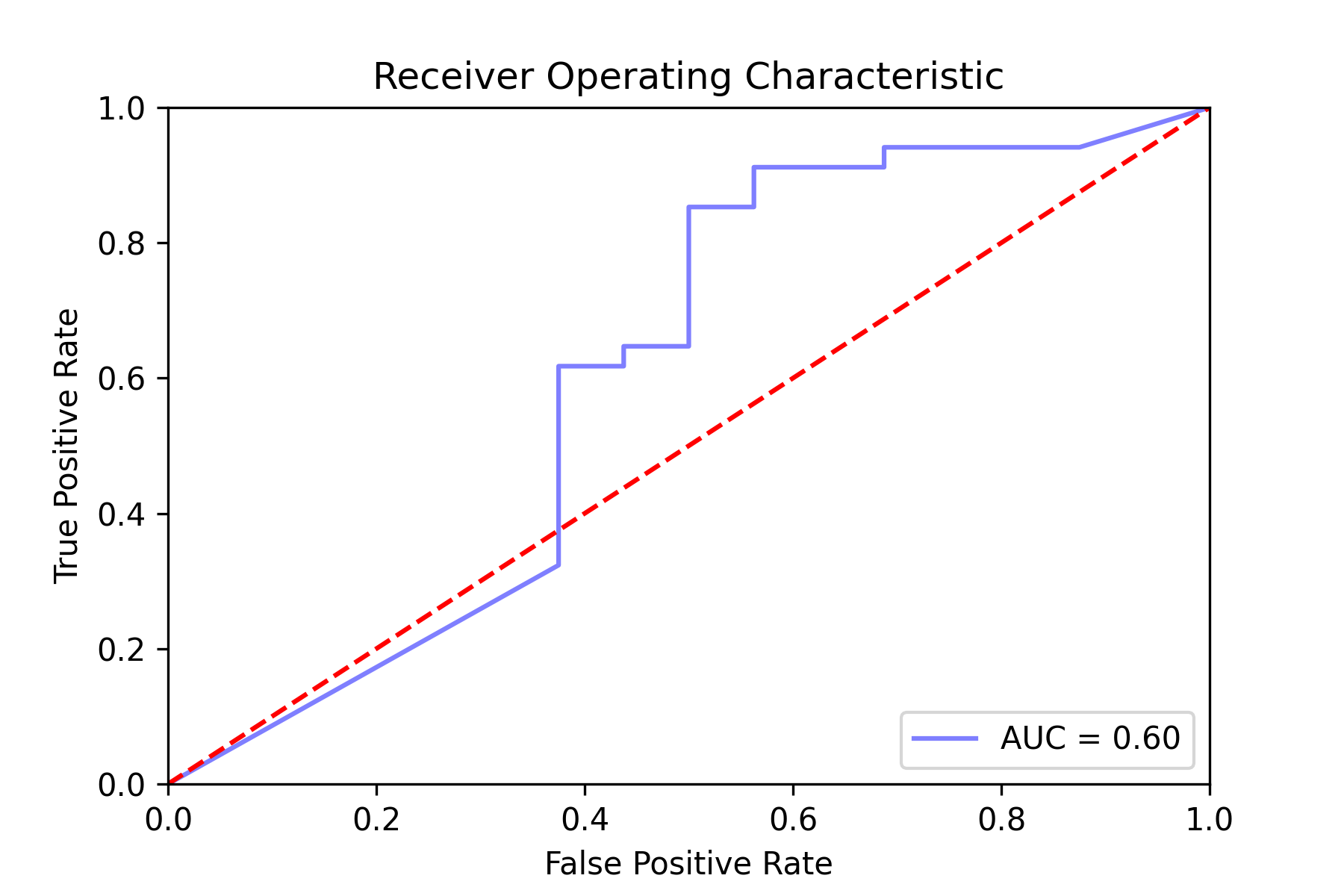} 
\caption{{\bf ROC curve for the broad market model}. 
We plot the ROC (Receiver Operating Characteristic) 
curve for the FactSet Us Index model, showing 
the true positive rate and the false positive rate, 
which summarize the performance of our probabilistic 
logit model in the validation set. 
The curve, being above the bisector most of the 
time, demonstrates that our model has a better 
than random level of accuracy of $60\%$, as measured 
by the AOC (Area Under the Curve). 
}
\label{fig:figure_roc_FactSet}
\end{figure}

\begin{figure}[h]
\includegraphics[width=0.9\textwidth]{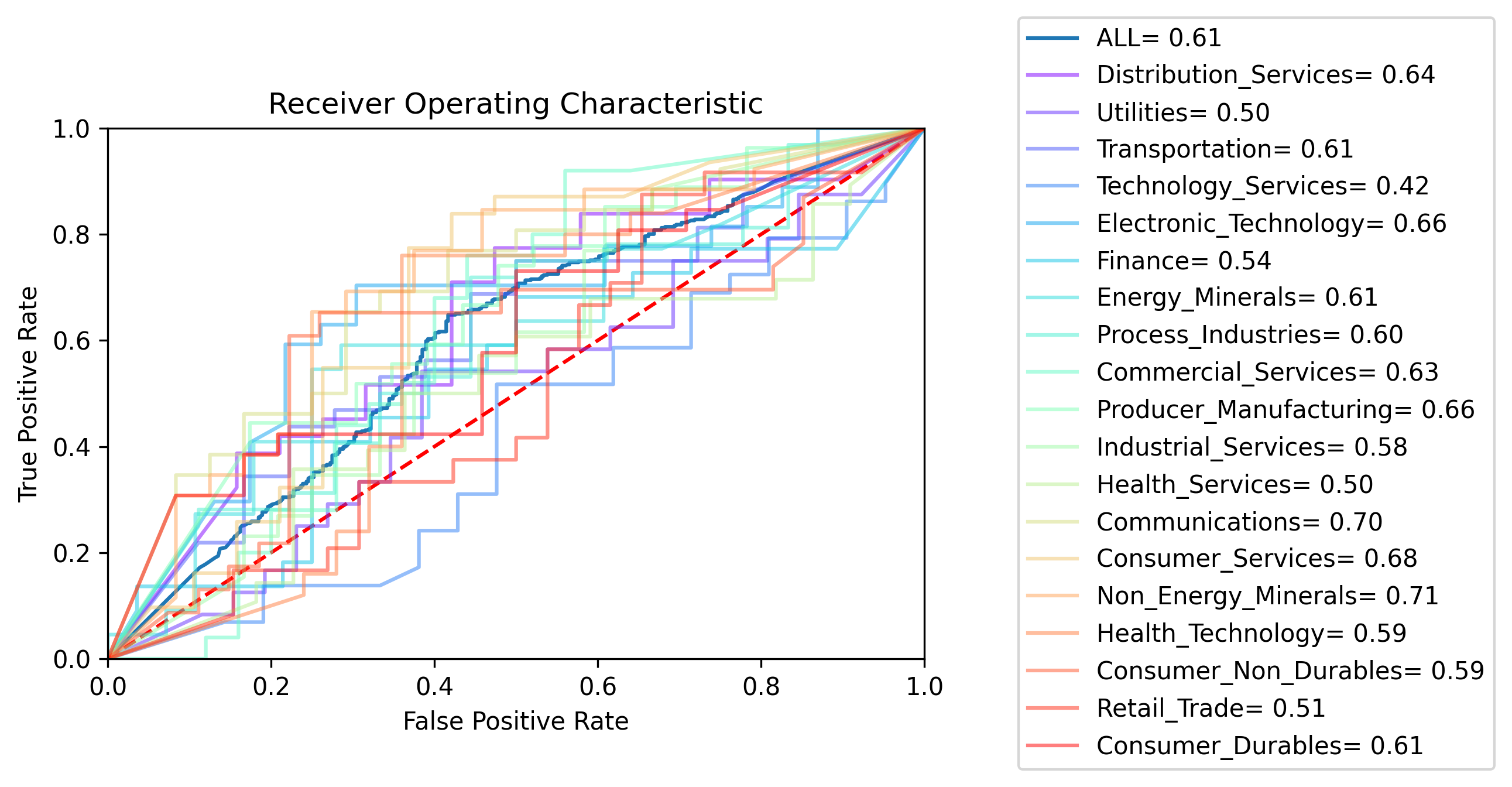} 
\minipage{0.32\textwidth}
  \includegraphics[width=\linewidth]{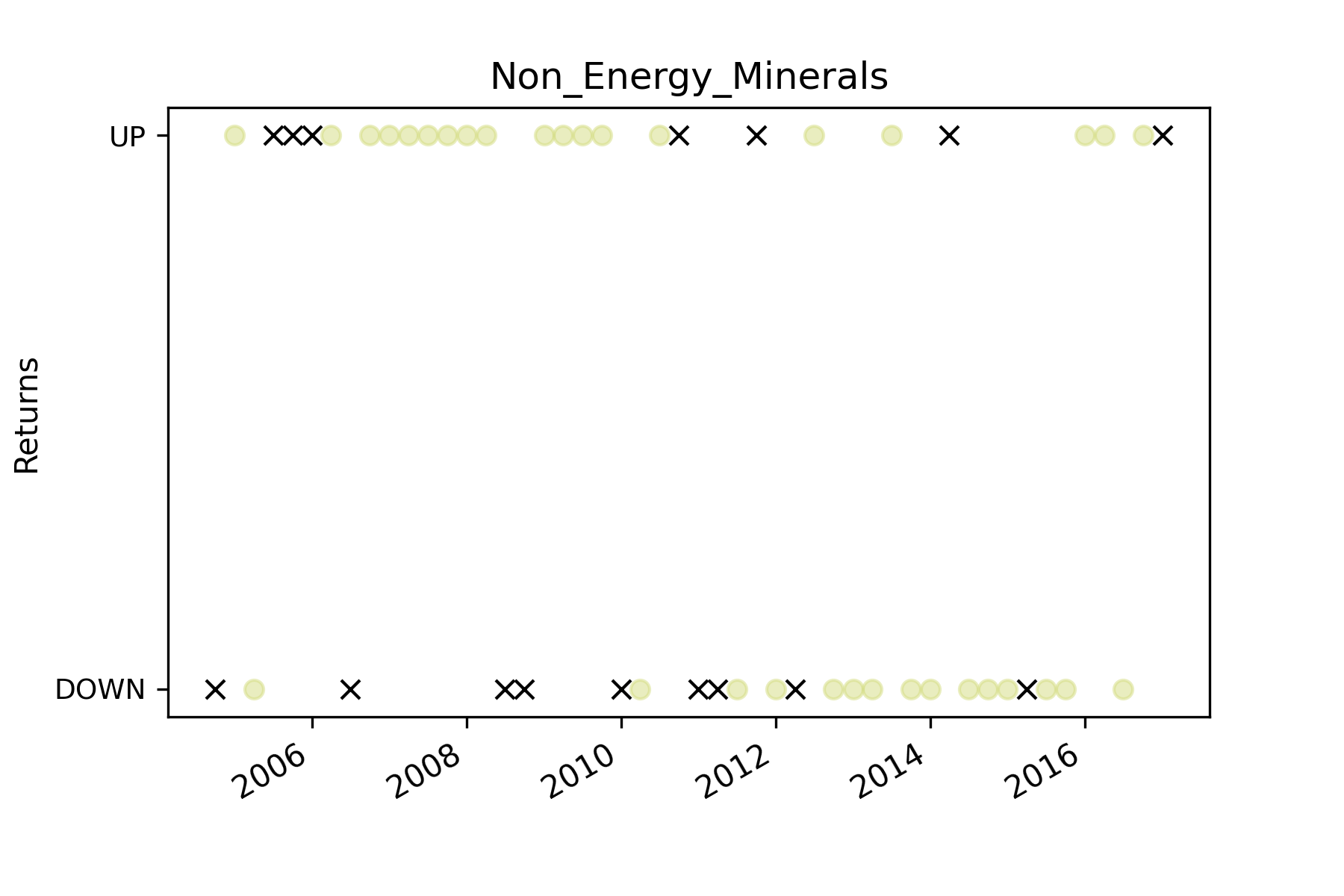}
\endminipage\hfill
\minipage{0.32\textwidth}
  \includegraphics[width=\linewidth]{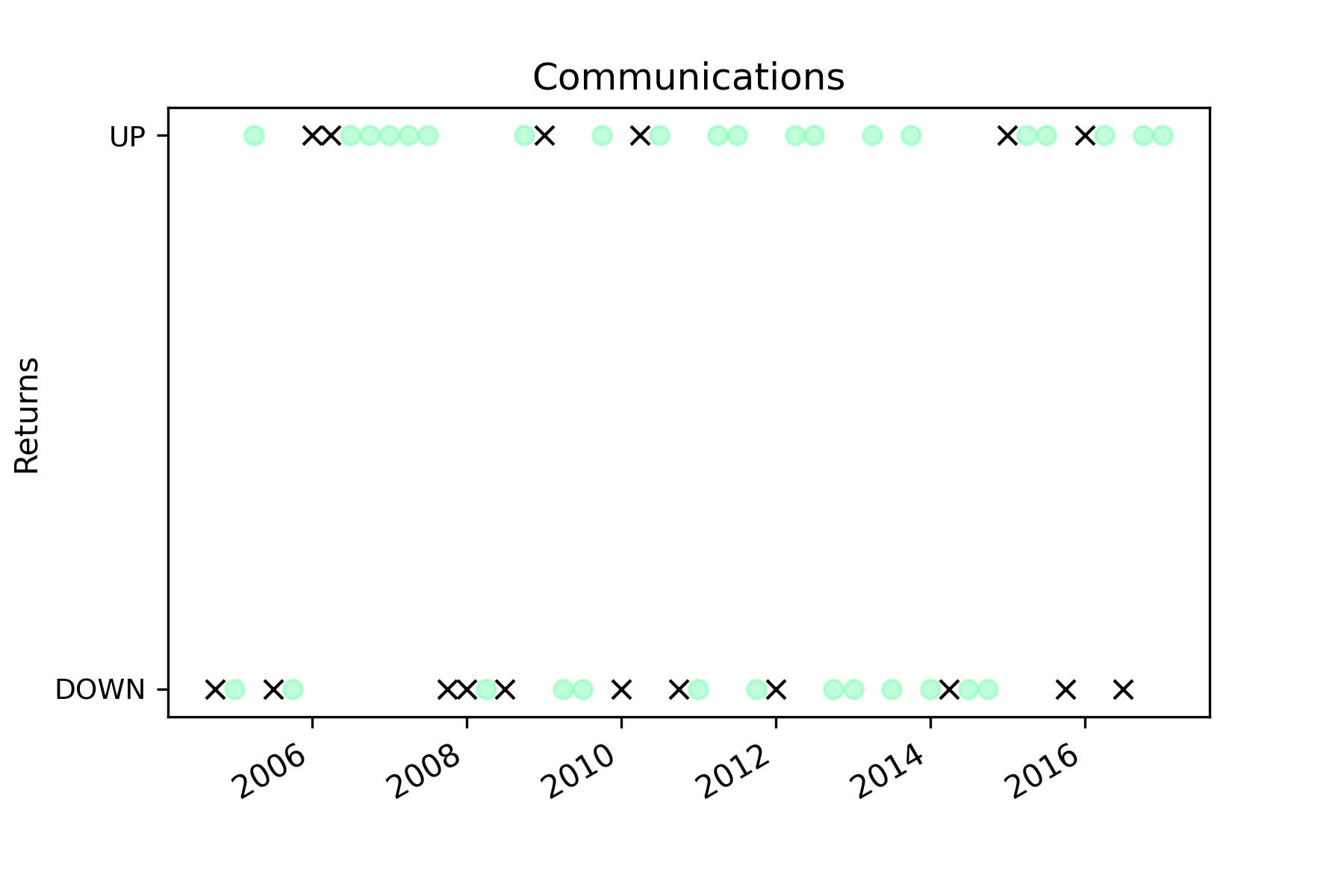}
\endminipage\hfill
\minipage{0.32\textwidth}%
  \includegraphics[width=\linewidth]{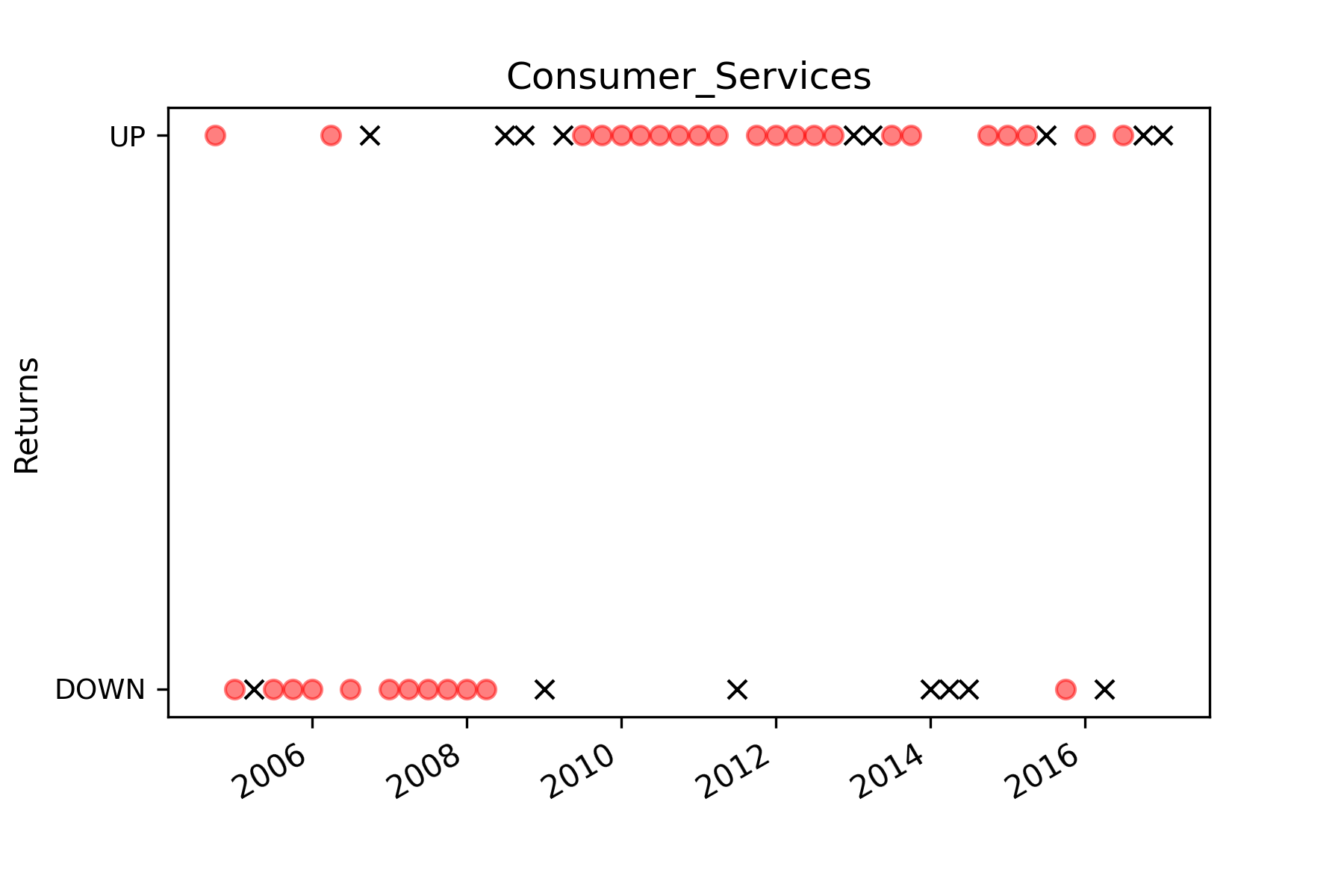}
\endminipage
\caption{{\bf (Top figure) ROC curves for the 
sector models.} 
We plot the ROC curve for each sector model, 
showing the true positive rates and the false 
positive rates. 
The predictive accuracy of the model goes from 
a maximum of $71\%$ for Non Energy Minerals to 
a minimum of $42\%$ for Technology Services. 
We also include the average curve obtained by 
aggregating the classification labels for all 
sectors (ALL), corresponding to an average 
accuracy of $61\%$. 
{\bf (Bottom figures)  UP/DOWN Return Predictions}
We plot our binary predictions vs the ground truth 
values for the sectors with the best performances 
in terms of AUC (Non Energy Minerals, Communications, 
and Consumer Services). Correct predictions are 
the dots and the incorrect ones are the crosses.
} 
\label{fig:figure_roc_sectors}
\end{figure}

We evaluate the accuracy of the estimated logit 
model by applying it to `evaluation' data where 
we know the correct output value. Evaluation 
data, as usual, has been withheld from model estimation 
to avoid overfitting.
Quantitatively, we measure the accuracy of the 
model by constructing the ROC (Receiver Operating Characteristic) curve. 
We include the results of two models: a model 
to predict the broad market, i.e. the FactSet 
US Index, which does not consider the sectors' 
information (Fig.~\ref{fig:figure_roc_FactSet}); 
and a sector-aware model which includes all 19 
market sectors (Fig.~\ref{fig:figure_roc_sectors}). 
ROC curves are constructed by comparing the 
probabilistic predictions of the logit model, 
$P({\rm UP})$, with a varying threshold $\theta$ 
to obtain classification labels, namely 
$c_i = \text{sign}[P({\rm UP})-\theta]$. 
The labels are then compared with the real labels 
in the validation set representing the sign of 
the returns for the corresponding broad market 
or sectors. 
By construction, when a curve lies consistently 
above the bisector then the $TPR$ (True Positive Rate) 
is above the $FPR$ (False Positive Rate), so 
the model is performing consistently better than 
a null model based on the simple label statistics. 

\begin{figure}[h]
\includegraphics[width=0.75\textwidth]{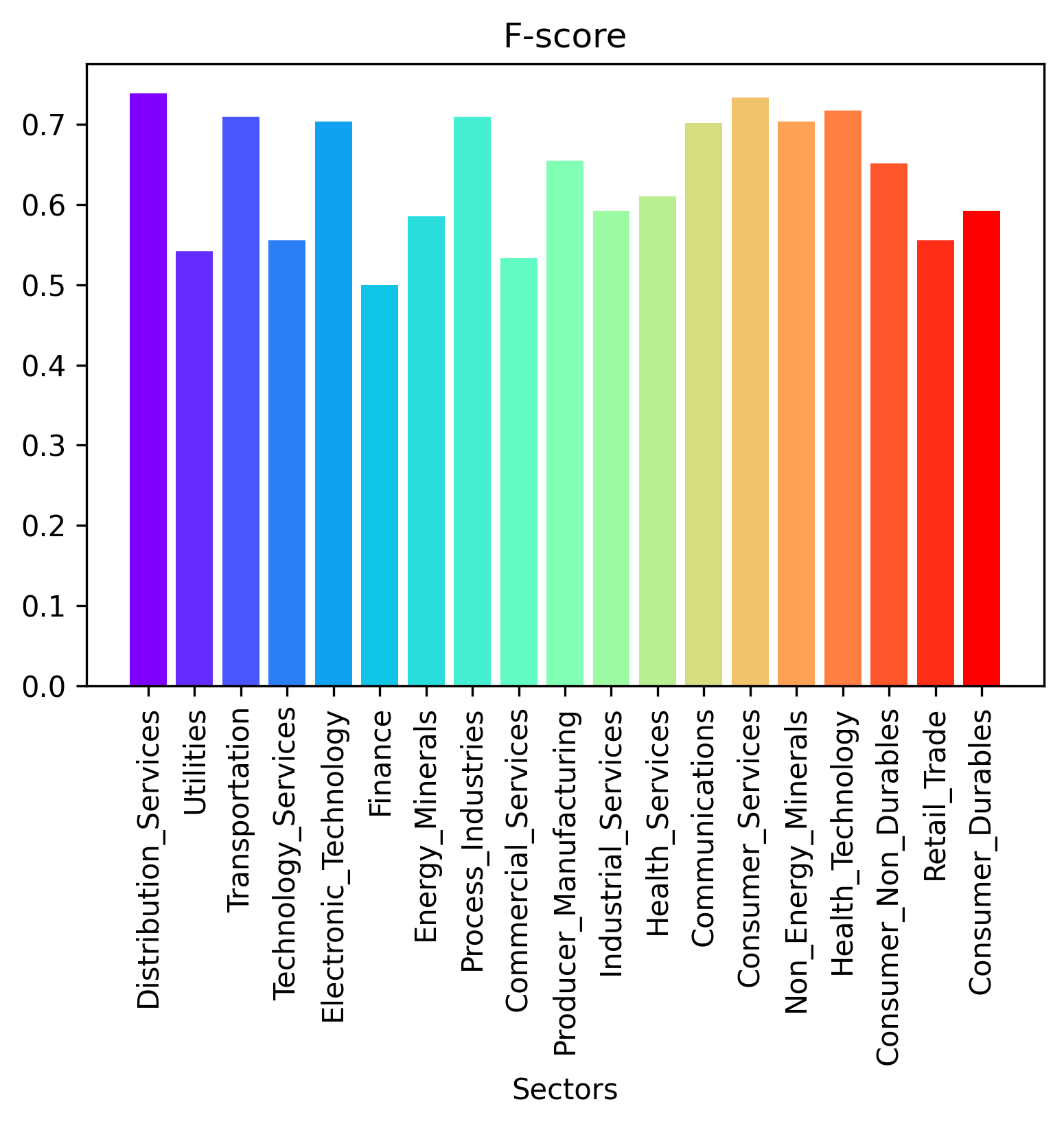} 
\caption{{\bf F1 score values for the Sector models.} 
We show the F1 values for each sector model, i.e. 
the ratio between the precision-recall product and their 
average. The average F1 score across sectors is $0.64$. 
} 
\label{fig:figure_f1_sectors}
\end{figure}

The result of the model are shown in Fig.~\ref{fig:figure_roc_FactSet} and Fig.~\ref{fig:figure_roc_sectors} for the 
FactSet US Index (broad market) and for the 
19 sectors comprising the FactSet Us Index, 
respectively.   
The information provided is enough to display 
a better than random level of accuracy, i.e. 
the ROC curve is above the bisector for most 
threshold values. 
An even better result is also found at a sector 
level, as seen in Fig.~\ref{fig:figure_roc_sectors}.  
In this case we find that for all sectors but 
{\it Utilities}, {\it Health Services}, and {\it Technology Services}, the ROC curves are 
consistently above the bisector. 
Furthermore, when aggregating the 
sector-level results we get a better accuracy 
than with the broad market model, 
in the sense that the average curve (labeled `ALL')
in Fig.~\ref{fig:figure_roc_sectors} lies 
above the bisector for all values of the abscissa, 
as opposed to the curve in Fig.~\ref{fig:figure_roc_FactSet}. 
This is because the model is able to account for 
the interplay of the sector dependency and the 
private investment information. 
Inspecting in more detail the predictions in the 
sectors with the best performances, i.e. {\it 
Non Energy Minerals}, {\it Communications} and 
{\it Consumer services}, we can observe the 
predominance of correct predictions over incorrect 
ones (dots over crosses), both true positives and true negatives, as well as their homogeneous distribution in the test set. For the best performing sectors 
accuracy reaches a maximum of $71\%$ as measured 
by the AUC (Area Under the Curve). 
Correct predictions are consistently found at all 
times with the exception of the {\it Technology Services} sector.  
To substantiate the assessment of the forecasting 
performance of our sector models, we considered also 
the F1 measure, 
shown in Fig.~\ref{fig:figure_f1_sectors}, and defined as 
the ratio between the product of precision and recall 
and their average according to the formula
\begin{equation}
F1 = 2\frac{PR}{P+R}\ .
\end{equation}
Notably, we find that the average F1 score across sectors 
is equal to 0.64, thus corroborating the forecasting performance 
evaluated from the ROC curve.  

\section{Discussion and Conclusions}
Information propagation is a key mechanism in finance. 
In this study, we have showed evidence of the signal 
that is present in private investments in companies 
and its role in affecting prices in public markets. 
We used data of private equity investments in private 
companies 
in a given quarter to predict the performance 
of the corresponding sectors in the public market in 
the following quarter. 
We have established via experiments that our hypothesis 
is correct, i.e. the presence of signals in these investments, 
as demonstrated by the performance of the logit 
models estimated on this information.

Nevertheless, we believe it is equally important 
to have the quantitative results supported by qualitative 
economic factors that explain them. 
Firstly, we observe that the percentage of sector counts 
to all deal counts seems like a rational choice consistent 
with our goal of predicting relative sector behavior (i.e.,
of a sector to a broad index). 
It is also reasonable that the positive signals generated 
from these first-deal counts are more accurate than negative 
signals, because there is some asymmetry between the decision 
of a fund manager to invest in a new company and to not invest 
or divest.  In fact, the decision to invest reflects an 
affirmative forecast by the portfolio manager, but the decision 
to not invest or divest may be more passive, i.e. driven by 
other factors such as dry powder, fund raising, the cost of capital, 
or profit taking.
However, such asymmetry may be less important for sectors 
heavily driven by commodities prices, such as all the 
{\it Energy Minerals, Non-Energy Minerals}, and {\it Utilities} 
sectors. 

Another important observation concerns the use of 
quarterly data rather than higher frequency data, 
such as daily or weekly data. 
The rationale behind our choice of quarterly data 
is that using daily or weekly data may be inferior 
to using quarterly data, as the signals may be too 
responsive to daily/weekly changes that are highly 
erratic. 
Furthermore, the horizon at which an investor may be 
privy to privileged information is important, as well. 
For example, the investor may rely on confidential sales 
information, which may identify trends expected to persist 
for at least a quarter or two.
However, this may be not necessarily true for sectors such as 
{\it Financial Services}, which may be causally driven 
by public market behavior that is far less predictable. 
In addition, the value of such companies is mostly driven 
by mark-to-market valuations and trading profits which 
are virtually impossible to predict.
At any rate, our forecasts are generally robust over 
time, i.e. correct predictions do not disappear 
over time and are found throughout the test set. 

Further studies will be devoted to investigating the 
forecasting performance of private equity signalling 
before and after the Covid-19 crisis, which represents 
an economic environment different from those contemplated 
in this work.
Eventually, an interesting future research direction would 
be the study of the profitability of a trading strategy 
based on this private investment signal as well as understanding 
the potential improvement of the model performance 
when including more concurring factors for the 
prediction as well as the statistical limits of 
the robustness displayed over time and across 
sectors. 

\clearpage

\noindent
{\bf \large Data availability}
Data that support the findings of this 
study are available at the FactSet 
database at~\url{https://www.factset.com/}.
Private Equity data are collected by FactSet. 
In this work we use a dataset from 01/01/2000 
to 12/31/2016.

\medskip

\noindent
{\bf \large Acknowledgments} 
This work was partially supported by AFOSR: 
Grant FA9550-21-1-0236. 

\medskip

\noindent
{\bf \large Author contributions}
Both authors contributed equally to the work presented in this paper.

\noindent
{\bf \large Competing interests} 
The authors declare no competing interests. Correspondence should 
be addressed to F.M. (fm2452@nyu.edu).

\section{List of abbreviations}
AUC: Area Under Curve, AUM: Assets under Management, FNR: False Negative Rate, FPR: False Positive Rate, PE: Private Equity, ROC: Receiver Operating Characteristic,  TPR: True Positive Rate, TNR: True Negative Rate. 


\bibliographystyle{unsrtnat}
\bibliography{pebib}

\end{document}